\documentclass[10pt,journal,compsoc]{IEEEtran}

\usepackage{amsmath,amsfonts}
\usepackage{algorithm}
\usepackage{algpseudocode}
\usepackage{array}
\usepackage[caption=false,font=normalsize,labelfont=sf,textfont=sf]{subfig}
\usepackage{textcomp}
\usepackage{stfloats}
\usepackage{url}
\usepackage{verbatim}
\usepackage{graphicx}
\def\BibTeX{{\rm B\kern-.05em{\sc i\kern-.025em b}\kern-.08em
    T\kern-.1667em\lower.7ex\hbox{E}\kern-.125emX}}
\usepackage{balance}
\usepackage{cite}  
\usepackage{listings}
\usepackage{multirow}
\usepackage{float} 
\usepackage{xspace}
\usepackage{makecell}

\usepackage{booktabs}
\usepackage[table,xcdraw]{xcolor}
\definecolor{dimgray}{gray}{0.95}

\renewcommand{\arraystretch}{1.5} 
\newcommand{\name}{IDEA\xspace}

\definecolor{aspgreen}{RGB}{0,128,0}
\definecolor{aspred}{RGB}{200,0,0}
\definecolor{aspblue}{RGB}{0,0,200}

\lstdefinelanguage{ASP}{
    morekeywords={hard_constraint,soft_positive_constraint,soft_negative_constraint,not,D1,D2,D3},
    keywordstyle=\color{aspblue},
    keywords=[2]{X,k,e11,e13,e21,e31},
    keywordstyle=[2]\color{aspgreen},
    sensitive=true,
    morecomment=[l]{\%},
    commentstyle=\color{gray},
    stringstyle=\color{aspred},
    basicstyle=\ttfamily\footnotesize,
    breaklines=true,
    postbreak=\mbox{\textcolor{aspblue}{$\hookrightarrow$}\space},
    columns=fullflexible
}

\newcommand{\inlineasp}[1]{\lstinline[language=ASP,basicstyle=\ttfamily\footnotesize\color{aspblue}]|#1|}

\begin{document}


\title{\name: Augmenting Design Intelligence through Design Space Exploration}

\author{
Chuer Chen,
Xiaoke Yan,
Xiaoyu Qi,
and~Nan Cao
\IEEEcompsocitemizethanks{
\IEEEcompsocthanksitem Chuer Chen is with the Intelligent Big Data Visualization Lab, Shanghai Research Institute for Intelligent Autonomous Systems, Tongji University. E-mail: chuerchen@tongji.edu.cn. 
\IEEEcompsocthanksitem Xiaoke Yan, Xiaoyu Qi, and Nan Cao are with Intelligent Big Data Visualization Lab, Tongji University. \\Email: \{2433547,qixiaoyu,nan.cao\}@tongji.edu.cn. \\Nan Cao is the corresponding author. 
}
}


\maketitle

\begin{abstract}
Design spaces serve as a conceptual framework that enables designers to explore feasible solutions through the selection and combination of design elements. However, effective decision-making remains heavily dependent on the designer's experience, and the absence of mathematical formalization prevents computational support for automated design processes. To bridge this gap, we introduce a structured representation that models design spaces with orthogonal dimensions and discrete selectable elements. Building on this model, we present \name, a decision-making framework for augmenting design intelligence through design space exploration to generate effective outcomes. Specifically, \name leverages large language models (LLMs) for constraint generation, incorporates a Monte Carlo Tree Search (MCTS) algorithm guided by these constraints to explore the design space efficiently, and instantiates abstract decisions into domain-specific implementations. We validate \name in two design scenarios: data-driven article composition and pictorial visualization generation, supported by example results, expert interviews, and a user study. The evaluation demonstrates the \name's adaptability across domains and its capability to produce superior design outcomes.
\end{abstract}


\section{Introduction}
\IEEEPARstart{A} design space serves as a structured repository of design knowledge that codifies design possibilities through interdependent dimensions and their valid element combinations. It has been widely adopted across design communities, such as visualization design~\cite{shu2020makes} and architectural design~\cite{halskov2021filtering}, to assist designers in making rational decisions. Recently, with the rise of generative AI, large-scale models have been increasingly used in design automation. However, their outputs often violate domain constraints or lack precise control over stylistic details~\cite{chen2024viseval}. The construction of effective design spaces has thus become pivotal — not merely as academic exercises, but as foundational infrastructure enabling reliable design automation.

In the visualization community, researchers have progressively formalized design spaces by analyzing design exemplars and abstracting recurring patterns into structured references. Prior research can be broadly categorized into two directions: guiding designers through spaces for aesthetic or expressive choices~\cite{chen2022vizbelle, lan2021kineticharts}, and supporting tool development by structuring dimensions around system requirements~\cite{cui2019text, zhu2021augmenting}. However, despite its widespread adoption as a conceptual tool, the construction and application of design spaces face three persistent challenges. First, traditional space construction predominantly relies on subjective expertise rather than systematic methodologies, resulting in ad-hoc frameworks that struggle to generalize across domains or evolving requirements. Second, existing spaces exhibit limited expressiveness in modeling intricate interdependencies among multidimensional design parameters, constraining their utility in complex scenarios requiring dynamic trade-off analysis. Third, the decision-making process remains manual, exacerbating cognitive load while stifling efficiency and innovative potential.

To address the above challenges, we propose a structured design space model that mathematically formalizes design spaces as executable decision structures with orthogonal dimensions and discrete elements, enabling systematic construction, constraint modeling, and automated decision-making. Building on this foundation, we introduce \name (i.e., \underline{i}ntelligent \underline{d}esign space \underline{e}xploration and \underline{a}ugmentation), a decision-making framework that leverages the design space to guide generative models, producing effective design outcomes from given requirements through three synergistic modules: (1) a LLM-powered constraint generation module that translates requirements into logical rules, (2) a constraint-guided Monte Carlo Tree Search that navigates the space for optimal solutions, and (3) an adaptive action execution module that synthesizes domain-specific implementations from abstract decisions. We demonstrate \name's cross-domain effectiveness in cases of data-driven article composition and pictorial visualization generation. The major contributions of the paper are as follows:

\begin{itemize}
    \item We propose a structured design space model that formalizes design knowledge into orthogonal dimensions and discrete elements as machine-interpretable decision structures, thereby enhancing computational understanding and intelligent design.

    \item We introduce \name, an intelligent design decision-making framework that transforms user requirements into high-quality outcomes through constraint-guided exploration of design spaces, with adaptable structures that support diverse domains.

    \item We validate \name’s effectiveness and generalizability through two scenarios: data-driven articles and pictorial visualizations, with two sets of example results, a user study confirming outcome quality, and expert interviews affirming human-aligned decision-making and practical applicability.
\end{itemize}

\section{Related Work}
In this section, we review the most relevant prior work across three areas, including design space, computational design, and algorithmic decision-making in design.

\subsection{Design Space}
The conceptualization of design spaces as structured frameworks has evolved across design disciplines. Mackinlay~\cite{mackinlay1986automating} first characterized it in the 1980s as "a space of possible designs". Recently, Halskov and Lundqvist~\cite{halskov2021filtering} refined this perspective to designer agency, defining it as "the space of design possibilities and alternatives that a designer considers to meet a design brief". Another influential perspective came from mathematical formalization. Shaw~\cite{shaw2011role} defined design spaces as "discrete Cartesian spaces where design decisions form orthogonal dimensions, alternatives constitute dimension values, and complete designs are space points". Similarly, Schulz et al.~\cite{schulz2010design} decomposed the concept into actionable parameters, proposing that "a design space comprises finite dimensions, each capturing a specific design decision required to fully specify a solution".

Building on these foundational definitions, visualization researchers have explored design spaces across various applications. In traditional visualization, structured design spaces have been developed for core techniques including scatterplots~\cite{sarikaya2017scatterplots}, maps~\cite{hografer2020map}, graph layouts~\cite{bludau2023unfolding,nobre2019state}, and matrix visualizations~\cite{rufiange2012treematrix}. The emergence of narrative visualization has expanded the design space to encompass more dimensions and choices, accommodating storytelling logic and structural coherence. For example, Bach et al.~\cite{bach2022dashboard} proposed an 8-dimensional dashboard design space, distinguishing content-oriented dimensions (e.g., data, visual representation) from composition-focused ones (e.g., page layout, color, structure). Yang et al.~\cite{yang2021design} analyzed data stories using Freytag's Pyramid~\cite{freytag1895technique} narrative structure, and identified a design space consisting of three dimensions: narrative pattern, data flow, and visual communication.

While existing design spaces provide structured references for human designers, their formulation lacks computational interpretability, preventing intelligent design. Building on Shaw's Cartesian model~\cite{shaw2011role} and Schulz's parametric decomposition~\cite{schulz2010design}, we propose a structured design space model that formalizes dimensions as orthogonal axes, elements as discrete nodes, and actions as executable functions, bridging the gap between conceptual exploration and automated implementation.

\subsection{Design Automation} Design emerges from a series of deliberate decisions, such as selecting color schemes, arranging visual hierarchies, and configuring typography. The first step toward design automation lies in understanding how these decisions collectively shape effective designs. For instance, Zhao et al.~\cite{zhao2018characterizes} proposed a deep ranking framework to learn a convolutional neural network for exploring the effects of various design factors (e.g., color, font) on graphic designs. Baechler et al.~\cite{baechler2024screenai} designed a vision-language model for UI and infographics understanding, including tasks such as question answering, element annotation, and summarization. 

To automate the complicated design creation process, researchers take a further step toward intelligent design generation. Early studies have investigated template-based methods. Cui et al.~\cite{cui2019text} developed an authoring system that automatically converts statements to a set of infographics with pre-designed styles. Yang et al.~\cite{yang2016automatic} proposed topic-dependent layout templates grounded in aesthetic principles and a computational framework to integrate layout elements under template constraints. Recent advance in multimodal foundation models enables high-quality generation by learning-based approaches. POSTA~\cite{chen2025posta} is a modular framework powered by diffusion models and multimodal large language models (MLLMs) for customized artistic poster generation. DesignDiffusion~\cite{wang2025designdiffusion} enables end-to-end text-to-design generation, jointly producing images and visual text elements. While these methods achieve unprecedented quality in specific domains, they require framework redesign and model retraining for new design problems. We address this limitation through \name — a universal framework combining decision-making algorithms and large foundation models to generate compelling designs, enabling customizations across diverse domains.

\subsection{Algorithmic Decision-Making in Design}
Decision-making algorithms optimize choices by defining reward functions and evaluating policies in dynamic environments. Core methods include heuristic approaches (e.g. decision trees~\cite{rokach2005decision}, genetic algorithms~\cite{goldberg1989genetic}) and learning-based methods (e.g. reinforcement learning~\cite{sutton1998reinforcement}), which iteratively refine decisions through experience, balancing short- and long-term rewards. Most design tasks involve sequential decision-making, where interdependent choices constrain future options. This temporal dependency increases complexity and calls for anticipating downstream impacts. Markov Decision Processes (MDPs)~\cite{puterman2014markov} formalize this through state-action-reward frameworks, enabling probabilistic modeling of decision cascades via value-based algorithms~\cite{watkins1992q, mnih2015human}, policy-based algorithms~\cite{williams1992simple, schulman2017proximal}, and Monte Carlo Tree Search (MCTS)~\cite{browne2012survey}. Recent advances integrate large language models (LLMs) as knowledge-aware planners~\cite{cao2024survey,yang2024selfgoal,tan2024true,ahn2022can,yan2023ask}, enhancing complex decision reasoning through pre-trained world knowledge.

Visualization design inherently constitutes a decision-making process involving sequential choices in data selection, transformation, and presentation. Emerging approaches embed algorithmic decision-making for intelligent generation. For instance, Xie et al.~\cite{xie2024haichart} implemented Monte Carlo Graph Search with composite rewards to automate operations like axis binding and mark selection. Shi et al.~\cite{shi2020calliope} developed logic-oriented MCTS to generate data-driven narratives by exploring facts in datasets. Shen et al.~\cite{shen2024data} proposed an LLM-based multi-agent framework where agents collaboratively decide on insights extraction, visual encoding, and narrative structuring for data video generation. These works highlight the potential of decision-making algorithms for smarter, more efficient design automation. Building on them, we introduce a unified decision model that integrates LLM-generated constraints into reward functions and employs MCTS-guided search for adaptable automation across diverse design spaces.

\section{Structured Design Space Model}
A design space is a conceptual framework that structures all feasible solutions to a design problem through explicit variables and their interdependencies. Early work~\cite{halskov2021filtering} framed it as a navigable realm of possibilities, while Shaw~\cite{shaw2011role} mathematically formalized it as a "discrete Cartesian space of orthogonal design dimensions" — a critical step toward computational thinking. Inspired by Schulz et al.’s~\cite{schulz2010design} and Shaw~\cite{shaw2011role}, we propose a structured design space model that bridges conceptual abstraction and algorithmic execution.

To formalize this model, let a design space $S$ be defined by $n$ design dimensions:

\begin{equation}
    S = \{D_1,D_2,...,D_i,...D_n\}
\end{equation}
Each $D_i$ is an orthogonal design dimension representing an independent aspect of the design problem, combinable to explore diverse solutions.

A design dimension establishes an orthogonal decision axis, with each design element as a selectable option contributing to the overall solution. Formally, each design dimension $D_i$ consists of a finite set of design elements:

\begin{equation}
    D_i = \{e_{i1},e_{i2},...,e_{ij},...e_{im_i}\}
\end{equation}
where $e_{ij}$ denotes the $j$-th element within the $i$-th dimension, and $m_i=|D_i|$ represents the number of available design elements in $D_i$.

Based on the proposed structure of the design space, a design solution is constructed by selecting one or more elements from each dimension. Formally, a design solution $P$ is defined as:

\begin{equation}
    P=\{E_1, E_2,...,E_i,...E_n\}
\end{equation}
where each $E_i \subseteq D_i$ represents the selected design elements from the $i$-th dimension. While most dimensions require one element to be selected ($|E_i|=1$), some support the selection of multiple elements ($|E_i|\geq1$). We denote these dimensions allowing multiple element selections as $\mathcal{D}_{multi} \subseteq S$.

Once a design solution $P$ is determined, it requires transformation into a final design outcome by implementing the selected design elements through concrete actions. To formalize this process, we defined a set of action functions $\mathcal{A}=\{A_1,...,A_m\}$ where each $A_j$ operates on elements from a specific subset of dimensions $\mathcal{D}_j \subseteq S$. The complete design outcome $O$ is generated through executing all actions over the design space:

\begin{equation}
O = \bigoplus_{j=1}^m A_j\left( \bigcup_{D_i \in \mathcal{D}_j} E_i\right)
\label{eq:action}
\end{equation}
where $A_j$ is the $j$-th action function, $\mathcal{D}_j$ denotes the subset of dimensions processed by $A_j$. $\bigoplus$ is a domain-specific composition operator that aggregates partial outcomes.

\begin{table}[t]
\centering
\begin{tabular}{l p{5.3cm}} 
\hline
\cellcolor{dimgray}\textbf{Design Dimensions} & \textbf{Design Elements} \\
\hline
\cellcolor{dimgray}\textbf{Headline} & Data highlighting, Stating an issue, Making an evaluation or judgment, Asking a question, Sensationalism \\
\cellcolor{dimgray}\textbf{Narrative Intent} & Inform, Explain, Persuade, Entertain \\
\cellcolor{dimgray}\textbf{Narrative Structure} & Inverted pyramid, Freytag pyramid, Drilling-down, Compositing \\
\cellcolor{dimgray}\textbf{$\text{Narrative Pattern}^+$} & Compare, Concretize, Repetition, Gradual reveal, Slow-down/Speed-up, Familiarization, Defamiliarization, Silent data, Physical metaphor, Humans behind the dots, Breaking the 4th wall, Rhetorical question, Call to action, Make a guess, Exploration, Addressing the audience, Convention breaking, Users find themselves \\
\cellcolor{dimgray}\textbf{Narrative Perspective} & First-person, Second-person, Third-person, Shifting perspective \\
\hline
\end{tabular}
\vspace{0.5em}
\caption{Overview of the narrative composition space.}
\label{tab:narrative}
\vspace{-2em}
\end{table}

To exemplify the model, consider a narrative composition space $S_{narrative}$ for data-driven articles. Drawing from studies on data storytelling~\cite{hao2024design,bach2018narrative,ojo2018patterns,wikinarrative}, we identify five core dimensions that structure the narrative, as shown in Table~\ref{tab:narrative}: Headline ($D_1$), Narrative Intent ($D_2$), Narrative Structure ($D_3$), Narrative Pattern ($D_4$), Narrative Perspective ($D_5$). Among these, Narrative Pattern is the dimension allowing multiple element selections ($\mathcal{D}_{multi}=\{D_4\}$), as different patterns can be applied within an article. A design solution $P$ specifies selections for each dimension. For example:
\begin{align*}
P = \big\{ &E_1=\{\text{"Stating an issue"}\}, \\
&E_2=\{\text{"Inform"}\}, \\
&E_3=\{\text{"Inverted Pyramid"}\}, \\
&E_4=\{\text{"Contrast"}, \text{"Concretize"}\}, \\
&E_5=\{\text{"Third-Person"}\} \big\}
\end{align*}
This $P$ would then be transformed into an article $O$ using the action function \texttt{generate\_article} ($A_1$), which generates both the article’s headline and content based on the selected elements from the five dimensions ($\mathcal{D}_1 = S_{narrative}$). The outcome $O$ is generated as:
\begin{equation*}
O = A_1\left( \bigcup_{D_i \in S_{narrative}} E_i \right)
\end{equation*}

By explicitly formulating these components, the proposed design space model provides a structured approach to navigating decisions, constructing solutions, and translating them into outcomes. This formulation bridges conceptual design exploration with computational implementation and lays a foundation for intelligent design generation.

\section{Design Decision-Making Framework}
Building upon the design space model, we propose \name, a design decision-making framework that generates optimal design outcomes based on given requirements. In this section, we first provide an overview of \name, then dissect its three core modules, followed by an evaluation of the search algorithm’s performance.

\subsection{Framework Overview}
\begin{figure*}[th]
  \centering
  \includegraphics[width=0.99\textwidth]{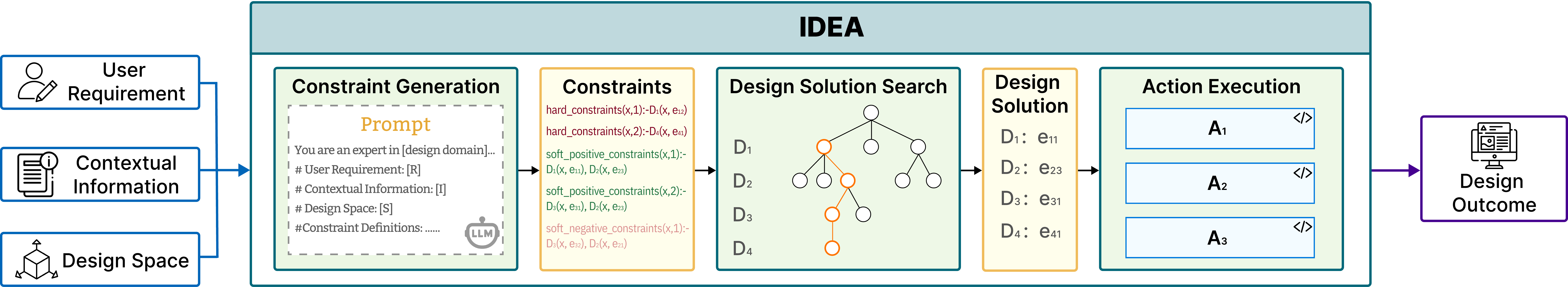}
  \caption{The overview of \name, including three core modules: constraint generation, design solution search, and action execution.}
  \label{fig:framework}
  \vspace{-1em}
\end{figure*}

\name is a versatile framework applicable across various design domains, providing a structured approach to decision-making for generating effective outcomes (Fig.~\ref{fig:framework}). For any design domain, \name operates based on three key inputs: user requirement $R$, design space $S$, and contextual information $I$. The user requirement defines the objective the generated design should fulfill, ensuring alignment with the intended goals. The design space provides a set of feasible options, establishing boundaries within which solutions can be explored. Contextual information supplies essential background data necessary for the design process, varying by domain—for instance, a visualization task requires a dataset, while interior design depends on room layout.

Upon receiving the inputs, \name first guides the LLM to generate design-space-specific constraints. These constraints are formulated with the user requirement as the objective and contextual information as a reference, steering the subsequent search for design solutions. Next, \name applies the Monte Carlo Tree Search (MCTS) algorithm~\cite{browne2012survey}, guided by a constraint-based reward function, to explore the design space and identify an optimal solution. Finally, it executes the corresponding action functions to generate the final design outcome that meets the user requirement.

\subsection{Constraint Generation}
The constraint generation module uses an LLM to formulate constraints that guide the search within the design space. Leveraging the LLM’s contextual reasoning and knowledge synthesis, the module takes the user requirement $R$, contextual information $I$, and design space $S$ as inputs, and outputs a set of constraints $C$ that both align with the requirement and remain grounded in the design space:

\begin{equation}
    C=LLM(R,I,S)
\end{equation}

\subsubsection{Constraint Definition}
To effectively guide the search within the design space, we adopt Answer Set Programming (ASP)~\cite{lifschitz2019answer} as the formalism for constraint representation, leveraging its declarative nature to specify constraints through logical rules rather than procedural algorithms, thereby ensuring they are both clearly defined and computationally enforceable.

The building blocks of ASP programs comprise \textit{atoms}, \textit{literals} and \textit{rules}. Atoms are elementary propositions with binary truth values. Literals are atoms or their negation with \texttt{not}. Rules establish logical dependencies through the structure $a \mathrel{:}{-}\,\,l_1,...,l_i.$ where the head atom $a$ is derived if all body literals $l_1,...,l_i$ are true.

In \name, design constraints are uniformly encoded as ASP rules with the structure $head \mathrel{:}{-}\,\, body.$, where the predicate name in the head explicitly identifies the constraint type. To ensure precise search control, we define three types of constraints: \textit{hard constraints}, \textit{soft positive constraints}, and \textit{soft negative constraints}, exemplified as follows:

\begin{lstlisting}[language=ASP,label=lst:asp-constraints]
hard_constraint(x, k) :- D1(x, e11).
soft_positive_constraint(x, k) :- D1(x, e13), D2(x, e21).
soft_negative_constraint(x, k) :- D3(x, e31).
\end{lstlisting}

Here, $x$ indicates the design instance being constrained, while $k$ is the index of the specific constraint. The body of the rule defines the conditions for the constraint, where each literal corresponds to a dimension (e.g., $D_1$) and its selected design element (e.g., $e_{11}$). 

Hard constraints enforce absolute prohibitions by defining conditions that invalidate a design solution when triggered. For example, in the rule \inlineasp{hard_constraint(x,k) :- D1(x,e11).}, the inclusion of $e_{11}$ in $D_1$ is prohibited, and the design is invalid if this condition is met. By explicitly defining forbidden regions in the design space, hard constraints enable efficient pruning of invalid solutions during search.

Soft positive constraints represent recommendations that improve design alignment with user requirements, not mandatory but desirable for better outcomes. For instance, the rule \inlineasp{soft_positive_constraint(x, k) :- D1(x, e13), D2(x, e21).} promotes the co-occurrence of $e_{13}$ in $D_1$ with $e_{21}$ in $D_2$. Such constraints establish preferential pathways through the design space, focusing the search on high-potential solutions while maintaining exploration flexibility. 

Conversely, soft negative constraints identify suboptimal patterns that should be minimized while remaining permissible. For example, the rule \inlineasp{soft_negative_constraint(x, k) :- D3(x, e31).} signals that $e_{31}$ in $D_3$ may reduce design effectiveness. Such constraints guide the search away from $e_{31}$ while permitting exceptional cases, maintaining solution diversity without over-restricting the design space.

\subsubsection{Prompt Design}
Design constraints exhibit high domain-specificity and user dependency, making static constraint templates ineffective across diverse scenarios. To address this, \name employs large language models (LLMs) as dynamic constraint generators, leveraging their cross-domain knowledge and contextual reasoning through an adaptable prompt framework (see supplemental materials) with five core components: \textit{role specification} (domain expertise definition), \textit{variable injection} ($R$,$I$,$S$), \textit{constraint semantics} (ASP syntax standards), \textit{generation rules} (quality control), and \textit{domain-anchored examples}. The constraint semantics provide reusable definitions, while variable injection dynamically feeds domain-specific inputs into constraint logic. Adapting \name to new domains requires only three adjustments: contextualizing role specifications, extending rules with domain rules, and supplying formatted examples. This structured customization preserves core logic while ensuring flexibility to accommodate domain-specific needs.

\subsection{Design Solution Search}
Guided by the generated constraints, which encode both design prohibitions and preferences, the framework employs Monte Carlo Tree Search (MCTS) to efficiently navigate the structured design space through iterative decision sequencing to find proper design solutions. In particular, the algorithm dynamically constructs a search tree $\mathcal{T}$ where each node $v$ represents a design element $e_{ij}$ from dimension $D_i$ and edges encode valid transitions between element selections. Each root-to-leaf path embodies a complete design solution $P$, generated through four cyclical phases: 

\begin{enumerate}
    \item \textbf{Selection} prioritizes nodes with highest constraint-guided upper confidence bound for trees (UCT).
    \item \textbf{Expansion} adds compliant child nodes corresponding to unexplored design elements.
    \item \textbf{Simulation} generates a complete solution via random walks and estimates its reward.
    \item \textbf{Backpropagation} updates node statistics (visit counts and rewards) along the traversal path.
\end{enumerate}


\begin{algorithm}[t]
\caption{Design Solution Search based on MCTS}
\label{alg:mcts}
\begin{algorithmic}[1]
\Require $S$, $C$, $W$, $\epsilon$
\Ensure $P^*$
\State $v_{root} \leftarrow \text{Node}(state=\emptyset)$ 
\State $\Delta_h \leftarrow [\,]$  
\State $converged \leftarrow \text{False}$  
\While{$\Delta<r_{max}$ \textbf{and not} $converged$}
    \Statex \hspace{\algorithmicindent} 
    \textcolor{olive}{\textit{\parbox[t]{\dimexpr\linewidth-\algorithmicindent}{\hangindent=1em // Traverse from root by selecting children with max UCT until reaching an expandable node $v$}}}  
    \State $v \leftarrow \text{Select}(v_{root},\mathcal{T})$ 
    
    \Statex \hspace{\algorithmicindent} 
    \textcolor{olive}{\textit{\parbox[t]{\dimexpr\linewidth-\algorithmicindent}{\hangindent=1em // Expand the current node $v$ by adding a new child node $v'$ for an unexplored design element in $S$}}} 
    \State $v' \leftarrow \text{Expand}(v, S)$ 

    \Statex \hspace{\algorithmicindent} 
    \textcolor{olive}{\textit{\parbox[t]{\dimexpr\linewidth-\algorithmicindent}{\hangindent=1em // Simulate a complete path from the new node $v'$ and evaluate its reward under constraints}}} 
    \vspace{1pt} 
    \State $\Delta \leftarrow \text{Simulation}(v',C)$ 
    
    \Statex \hspace{\algorithmicindent} 
    \textcolor{olive}{\textit{\parbox[t]{\dimexpr\linewidth-\algorithmicindent}{\hangindent=1em // Propagate reward back through the path to update node statistics in $\mathcal{T}$}}} 
    \vspace{1pt} 
    \State $\text{Backpropagate}(\mathcal{T},\Delta)$

    \Statex \hspace{\algorithmicindent} 
    \textcolor{olive}{\textit{\parbox[t]{\dimexpr\linewidth-\algorithmicindent}{\hangindent=1em // Check if rewards in the window have converged}}}  
    \State $\Delta_h.\text{append}(\Delta)$  
    \If{$\text{len}(\Delta_h) > W$}  
        \State $\Delta_h.\text{pop}(0)$ 
        \State $range \leftarrow \max(\Delta_h) - \min(\Delta_h)$  
        \State $converged \leftarrow (range \leq \epsilon)$  
    \EndIf  
\EndWhile
\Statex 
\textcolor{olive}{\textit{// Select the path with the highest reward as optimal design solution }} 
\State $P^* \leftarrow \text{BestPath}(v_{root},\mathcal{T})$
\State \Return $P^*$
\end{algorithmic}
\end{algorithm}

\textbf{Reward Function.} We propose a reward function to assess how well a design solution meets user requirements by calculating a constraint-aware reward and a selection quantity consistency penalty. 

First, we use an ASP solver~\cite{gebser2014clingo} to evaluate design solution $P$ against the generated constraints, obtaining counts of satisfied hard $V_h$, soft positive $V_p$, and soft positive constraints $V_n$. For hard constraints $C_h$ and soft negative constraints $C_n$, satisfaction indicates a violation of requirements, contributing negatively to the reward. In contrast, satisfying soft positive constraints $C_p$ suggests recommended elements are adopted, thereby increasing the reward. The constraint-aware reward is defined as:

\begin{equation}
    r_c(P) = -\alpha \cdot \frac{V_h}{|C_h|} + \beta \cdot \frac{V_p}{|C_p|} - \gamma \cdot \frac{V_n}{|C_n|}
\end{equation}
where $|C_h|$, $|C_p|$ and $|C_n|$ denote the total counts of each constraint type. $\alpha$, $\beta$ and $\gamma$ are weights that control their impact on the reward. To ensure hard constraints are strictly avoided and soft positive constraints effectively guide the search toward optimal design decisions, we set $\alpha=20$, $\beta=10$, and $\gamma=1$ after empirically tuning.

Next, for the dimensions allowing multiple element selections $\mathcal{D}_{multi}$, we enforce quantitative consistency between the selected count and the recommended count derived from soft positive constraints. Consider a scenario where soft positive constraints recommend three filters for the subspace dimension of data facts: if a solution deploys five filters (two over recommendation), this deviation triggers a linear penalty. The penalty function aggregates deviations across all dimensions in $\mathcal{D}_{multi}$ as follows:

\begin{equation}
r_q(P)= \sum_{D_k \in \mathcal{D}_{multi}} \big| |E_k| - n_k^{rec} \big|
\end{equation}
where $|E_k|$ denotes the number of elements selected in dimension $D_k$, and $n_k^{rec}$ is the recommended element count for that dimension derived from soft positive constraints.

Finally, the reward function is formulated as:
\begin{equation}
\label{eq:reward}
    reward(P)=r_c(P)+\delta \cdot r_q(P)
\end{equation}
where $\delta$ is the penalty weight, set to 0.5 in \name.

Based on the reward function, the constraint-guided upper confidence bound for trees (UCT) metric~\cite{kocsis2006bandit} balances the exploitation of high-reward solutions and exploration of less-visited elements. The UCT for a node $v$ is calculated as:

\begin{equation}
\text{UCT}(v) = \frac{r(v)}{n(v)} + c \cdot \sqrt{\frac{\ln n(parent(v))}{n(v)}} 
\end{equation}
where $r(v)$ is the reward of the node, $n(v)$ is the total visits to the node, $n(parent(v))$ is the visit count of its parent node, and $c$ is the tradeoff coefficient, which we set to $c=5$ after experiments.

\textbf{Algorithm Overview} The search algorithm (Alg.~\ref{alg:mcts}) constructs an optimal design solution through iterative exploration of the structured design space $S$ guided by generated constraints $C$. Initializing a search tree $\mathcal{T}$ with an empty root node, the algorithm iterates four phases: \textbf{Selection} traverses $\mathcal{T}$ by choosing child nodes maximizing the constraint-guided UCT metric; During \textbf{expansion}, when reaching a node $v$ representing element $e_{ij}$ from dimension $D_i$, the algorithm progresses to the next dimension $D_{i+1}$ by adding an unexplored element from $D_{i+1}$ as $v'$, while for dimensions allowing multiple element selections ($\mathcal{D}_{multi}$), it selects one remaining element within $D_i$ as $v'$ until reaching the maximum count or termination symbol (\textit{"none"}) before advancing. From the newly expanded node $v'$, \textbf{simulation} generates a complete solution $P$ evaluated by the reward function (Eq.~\ref{eq:reward}), with rewards \textbf{backpropagated} to update node statistics. The search terminates immediately upon finding a solution achieving maximum reward $r_{max}=\beta$ (when $V_h{=}0$, $V_n{=}0$, $V_p{=}|C_p|$, $r_q{=}0$), or if $r_{max}$ is unattainable due to conflicts (e.g., multiple soft positive constraints on the same dimension), once rewards stabilize within $W$ consecutive iterations under threshold $\epsilon$. Finally, the highest-reward path in $\mathcal{T}$ is extracted as $P^*$.

\subsection{Action Execution}
Building on the optimal design solution $P^*$, the action execution module realizes the formal model (Equation~\ref{eq:action}) by converting $P^*$ into a concrete outcome $O$ via customizable action functions, which developers implement for their domain. To preserve systematic execution, we establish an extensible pipeline that separates domain-specific action functions from the core logic while ensuring consistent execution patterns across domains.

Extensibility is achieved through a modular structure where developers define domain-specific actions and associate them with relevant design dimensions via a dimension-to-action mapping $\quad M_d: \mathcal{D}_j \mapsto A_j$. For instance, in data visualization, \texttt{x field} and \texttt{y field} map to a data encoding action, while \texttt{title type} directs to a title generation action. Given a design solution $P$, the module first aggregates elements by their associated actions, gathering elements from dimensions in $\mathcal{D}_j$ into a set $\mathcal{E}_j=\bigcup_{D_i \in \mathcal{D}_j} E_i$. Each action $A_j$ processes its assigned element set to produce partial outcomes $o_j=A_j(\mathcal{E}_j)$. The final outcome $O$ is synthesized from all these results through domain-specific rules, such as layering visual marks with text in visualization. By separating domain-specific action logic from generic execution mechanics, the framework ensures both adaptability and reproducibility.



\begin{table}[t]
\centering
\small 
\setlength{\tabcolsep}{2.5pt} 
\renewcommand{\arraystretch}{1.05} 
\begin{tabular}{lccccc}
\toprule
 & Best & Best & Conv. & Conv. & Validity \\
 & Reward ↑ & Time(s) ↓ & Reward ↑ & Time(s) ↓ & \textbf{Ratio ↑} \\
\midrule
\textbf{MCTS} & \textbf{9.692} & 9.680 & \textbf{9.692} & 18.980 & \textbf{0.980} \\
GA            & 9.376 & 4.969 & 9.351 & 20.597 & 0.978 \\
SA            & 8.737 & 6.659 & 7.931 & \textbf{6.765} & 0.628 \\
BS            & 8.556 & \textbf{1.026} & --     & --     & -- \\
\bottomrule
\end{tabular}
\vspace{0.5em}
\caption{Results of the quantitative comparison.}
\label{tab:quantitative}
\vspace{-2em}
\end{table}

\subsection{Evaluation of the Search Algorithm}
We quantitatively assess the performance of \name’s constraint-guided Monte Carlo Tree Search (MCTS) against three alternatives: genetic algorithms (GA), simulated annealing (SA), and beam search (BS). Reinforcement learning is excluded due to its need for retraining when constraints change, making it impractical for dynamic scenarios.

\textbf{Procedure.} We evaluated search algorithms within a data fact design space derived from the 2016 Rio Olympics medal dataset (313 rows, 6 columns). Ten distinct constraint sets were generated from this space, where each set admits an optimal solution achieving the maximum reward of 10. For each algorithm and constraint set, we executed 10 independent runs: MCTS, GA(population size=100), and SA(initial temperature=1000, cooling rate=0.999) terminated when reward fluctuation within a 100-iteration window fell below 0.1, while BS (beam width=5) completed full search trajectories due to lacking inherent convergence. We computed the average over 10 runs for each of the metrics: (1) best reward achieved during search and (2) time to reach this reward. For MCTS/GA/SA, we additionally measured three convergence-phase metrics: (a) time to meet convergence, (b) reward at convergence, and (c) valid solution ratio during search. BS metrics derive solely from its final search state, as convergence tracking was inapplicable (N/A).

\textbf{Results.} As shown in Table~\ref{tab:quantitative}, MCTS demonstrates superior solution quality and stability, achieving near-optimal average rewards (9.692/10) in both best and convergence phases, along with the highest validity ratio (98\%). This advantage stems from its exploration-exploitation balance, enabling consistent navigation toward high-reward regions. While BS achieves faster best results (1.03s) through deterministic path extensions, its greedy strategy confines the search to local optima, resulting in a lower maximum reward (8.556). SA converges quickly (6.76s) but struggles with suboptimal early solutions and limited global exploration, leading to a lower validity ratio and lower rewards. GA shows comparable stability but slower convergence and lower rewards than MCTS. The results validate MCTS as the optimal search algorithm for \name, balancing quality, efficiency, and constraint compliance.

\section{Case I: Data-Driven Article}
This section validates \name's effectiveness through its application to data-driven article composition. We develop an automated article composition system based on \name, operating within two structured design spaces. We first introduce the design spaces and system implementation, followed by a comprehensive evaluation combining example results and expert interviews.

\subsection{Design Spaces of Data-Driven Articles}
Data-driven articles combine statistics, visualizations, and narratives to convey persuasive insights. Their creation involves a series of design decisions, including following genre conventions, creating visualizations, and emphasizing key insights. To reduce the complexity of decision-making, we define two structured design spaces: one for textual components and another for data visualizations.

\begin{table}[t]
\centering
\begin{tabular}{l p{5.8cm}} 
\hline
\textbf{\makecell[l]{Design\\Dimensions}} & \textbf{Design Elements} \\
\hline
\cellcolor{dimgray} \textbf{Fact Type} & Value, Difference, Proportion, Trend, Categorization, Distribution, Rank, Association, Extreme, Outlier \\
\cellcolor{dimgray} \textbf{Breakdown} & Categorical or temporal fields, e.g., \textit{country}, \textit{city}, \textit{year} \\
\cellcolor{dimgray} \textbf{$\text{Measure}^+$} & Numerical fields with different aggregations, e.g., \textit{(sales, sum)}, \textit{(sales, maximum)} \\
\cellcolor{dimgray} \textbf{$\text{Subspace}^+$} & Filters combining categorical or temporal fields with specific values, e.g., \textit{(country, uk)} \\
\cellcolor{dimgray} \textbf{$\text{Focus}^+$} & Data items in the subspace, e.g. \textit{(year, 2024)}, \textit{(city, london)} \\
\cellcolor{dimgray} \textbf{\makecell[l]{Visualization\\Title}} & Data highlighting, Stating an issue, Making an evaluation or judgment, Asking a question \\
\hline
\end{tabular}
\vspace{0.5em}
\caption{Overview of the data fact space.}
\label{tab:datafact}
\vspace{-2em}
\end{table}

\begin{figure*}[th]
  \centering
  \includegraphics[width=\textwidth]{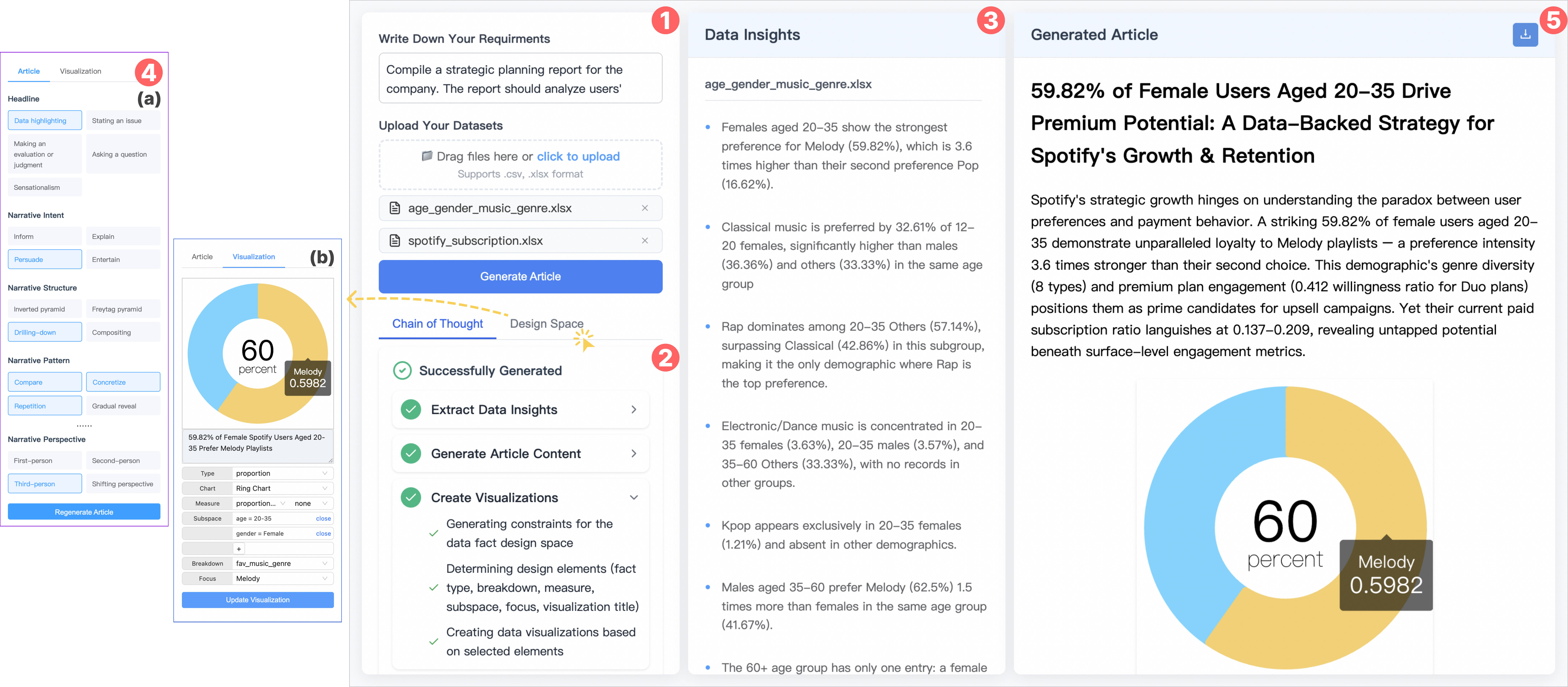}
  \caption{The user interface of the automated data-driven article composition system consists of five main components: (1) the input view, (2) the chain-of-thought view, (3) the data insights view, (4) the design space view, and (5) the article view.}
  \label{fig:interface}
    \vspace{-1em}
\end{figure*}

\textbf{Narrative Composition Space} The narrative composition space focuses on structuring the storytelling and guiding the communication of key insights through text. It consists of five dimensions previously identified in Table~\ref{tab:narrative}, which shape the narrative flow:

\begin{itemize}
    \item Headline: Determines how the article's headline expresses the core message.
    \item Narrative Intent: Defines the article's main purpose.
    \item Narrative Structure: Specifies information flow logic.
    \item $\text{Narrative Pattern}^+$: Represents a low-level narrative device for building the content. 
    \item Narrative Perspective: Refers to the point of view from which the article is told.
\end{itemize}

\textbf{Data Fact Space} The data fact space manages to extract the key information from the original data, enabling the generation of insightful data visualizations. Inspired by Shi~\cite{shi2020calliope} and Hao~\cite{hao2024design}, we formalize six core dimensions:

\begin{itemize}
    \item Fact type: Indicates the type of information described by the fact.
    \item Breakdown: Describes how the data is divided into subgroups or categories.
    \item $\text{Measure}^+$: Represents the quantifiable field or variable that is being measured.
    \item $\text{Subspace}^+$: Defines the specific subset of the data.
    \item $\text{Focus}^+$: Indicates the particular focus or point of emphasis within the subspace.
    \item Visualization Title: Determines how the key message or finding is communicated in the title.
\end{itemize}

As shown in Table~\ref{tab:datafact}, breakdown, measure, subspace, and focus dynamically instantiate from the dataset, while fact type and visualization title provide fixed elements. Notably, measure, subspace, and focus support multi-selection with tuple elements: measure uses (\textit{field}, \textit{aggregation}), subspace and focus use (\textit{field}, \textit{value}). This formalization effectively transforms data facts into a structured design space.

\subsection{Implementation Details}
We implement an automated data-driven article composition system using \name, which leverages DeepSeek-R1~\cite{guo2025deepseek} with chain-of-thought prompting for all LLM-based operations. The system begins by accepting a user’s high-level requirement (e.g., “research report”) and datasets. An LLM extracts multiple data insights from the uploaded datasets, which serve as contextual information for \name to generate article content and titles based on the narrative composition space. Subsequently, the system utilizes both the generated article content and original datasets as new contextual information, applying \name to the data fact space to create matching visualizations for paragraphs with strong data relevance. Finally, the textual and visual components are integrated into a complete data-driven article.

\textbf{User Interface.} To create a data-driven article, a user first needs to specify his requirement, upload datasets, and initiate generation through the input view (Fig.~\ref{fig:interface}-1). Once generation starts, the chain-of-thought view (Fig.~\ref{fig:interface}-2) displays the system’s step-by-step progress, providing transparency and helping users anticipate the output time and key stages of the process. Following data insights extraction, the processed insights for each dataset are displayed as lists in the data insights view (Fig.~\ref{fig:interface}-3). The system then determines optimal elements in the narrative composition space, accessible via tab navigation in the design space view (Fig.~\ref{fig:interface}-4(a)). The generated content appears in the article view (Fig.~\ref{fig:interface}-5), with generated visualizations automatically embedded at relevant positions. The user may click any visualization to edit its fact configurations in the visualization tab (Fig.~\ref{fig:interface}-4(b)). To modify the whole article, the user can switch to the article tab (Fig.~\ref{fig:interface}-4(a)) to edit selected elements, then trigger regeneration to update both textual content and associated visualizations simultaneously.

\begin{figure*}[th]
  \centering
  \includegraphics[width=0.95\textwidth]{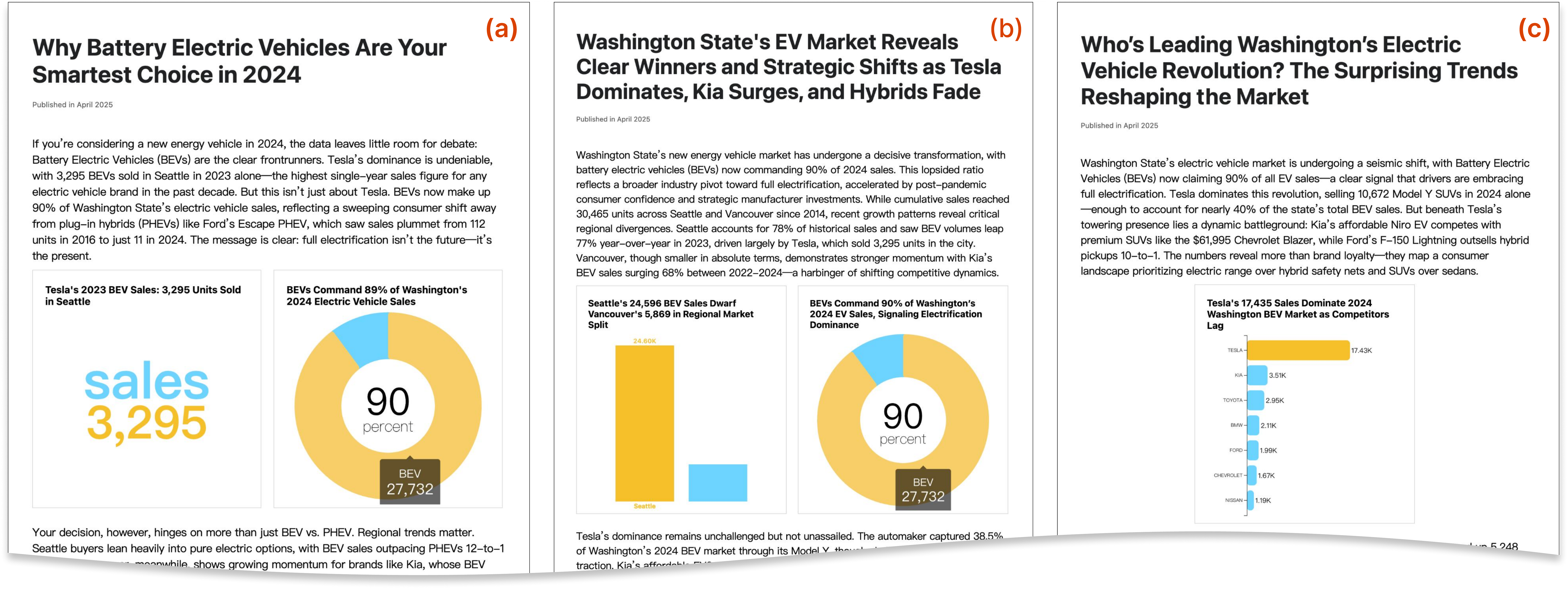}
  \caption{Data-driven articles generated by \name. (a) is a compelling buying guide about purchasing new energy vehicles; (b) is an in-depth industry report on the new energy vehicle market; (c) is an engaging automotive news article that explores the new energy vehicle market in Washington.}
  \label{fig:articles}
\end{figure*}

\subsection{Evaluation} We present three example results, followed by a series of interviews exploring system usability and decision quality.

\subsubsection{Example Results}
To validate \name's ability to make appropriate design decisions and produce high-quality outcomes, the system was used to create three types of data-driven articles under different user requirements as shown in Fig.~\ref{fig:articles} (see supplemental materials for full articles). The articles were generated based on two datasets: (1) electric vehicle (EV) sales (2014-2024) by brand for Seattle and Vancouver (221 rows, 5 columns), and (2) 2024 EV sales and prices by brand and model in Washington State (30 rows, 8 columns).

Fig.~\ref{fig:articles}(a) shows the article generated from the requirement: \textit{"A compelling buying guide about purchasing new energy vehicles, providing recommendations about type and brands for buyers. The article is designed for general consumers looking for practical advice."} Adopting the \textit{persuade} intent, the article employs a \textit{second-person} perspective and a (\textit{making an evaluation or judgment}) headline to engage readers. The \textit{inverted pyramid} structure delivers the core recommendation upfront, supporting the goal of offering clear, actionable advice.

Fig.\ref{fig:articles}(b) is generated based on the requirement: \textit{"An in-depth industry report on the new energy vehicle market, focusing on the top-selling brands in Washington State. The report should be tailored for professionals and investors."} The \textit{making an evaluation or judgment} headline sets an authoritative tone by synthesizing multi-dimensional trends into a strategic assessment suitable for investors. Adopting a \textit{third-person} perspective and \textit{explain} intent, the report follows a \textit{drilling-down} structure, aligning with the professionals’ needs for objective insights and analytical depth.

Fig.~\ref{fig:articles}(c) is generated from the requirement: \textit{"An engaging automotive news article aimed at automotive enthusiasts and general readers that explores the new energy vehicle market in Washington State, focusing on top-selling brands."} The headline adopts the \textit{asking a question} pattern to spark curiosity. Guided by the \textit{inform} intent, the article uses a \textit{third-person} perspective and \textit{inverted pyramid} structure to present key findings upfront. It maintains journalistic neutrality while ensuring accessible content for a broad audience.

\subsubsection{Expert Interview}
To further evaluate \name’s practicality and effectiveness, a series of interviews were conducted with 5 domain experts (E1-E5, 4 females, 1 male), each with an average of 4.8 years of experience in data storytelling. The interviews centered on our automated data-driven article composition system as an application of \name. 

\textbf{Baseline System.} To assess efficacy of \name's decision-making, a baseline system is developed where experts manually select elements from the narrative composition space. This baseline maintained identical action functions and UI components, with the sole variation being human selection versus \name's constraint-guided MCTS selection.

\textbf{Materials.} We prepared two datasets on storms in Miami over the past decade. Both systems were tested using the same requirement to compose a travel guide offering tourist advice and safety precautions for storm risks in Miami. A reference document explaining the narrative composition design space was provided to assist experts during manual selection in the baseline condition.

\textbf{Procedure.} The interviews were performed via an online meeting system with experts' consent to join and be recorded. Each interview began with a 10-minute introduction about the full and baseline system. Following this, experts were asked to use the baseline system to create a data-driven article by manually selecting elements from the narrative composition space based on their understanding. The system then generated article content and visualizations accordingly. Experts were encouraged to revise their choices or edit visualizations until satisfied, after which their final selections were recorded. Next, they used the full system with the same input, where \name automatically executed the entire pipeline. The auto-selected elements were also logged for comparison. Experts reviewed both versions side by side, followed by a semi-structured interview on their decisions and overall experience, and a questionnaire assessing system performance. Each interview lasted about one hour and was recorded for later analysis.

\begin{figure}[t]
  \centering
  \includegraphics[width=0.49\textwidth]{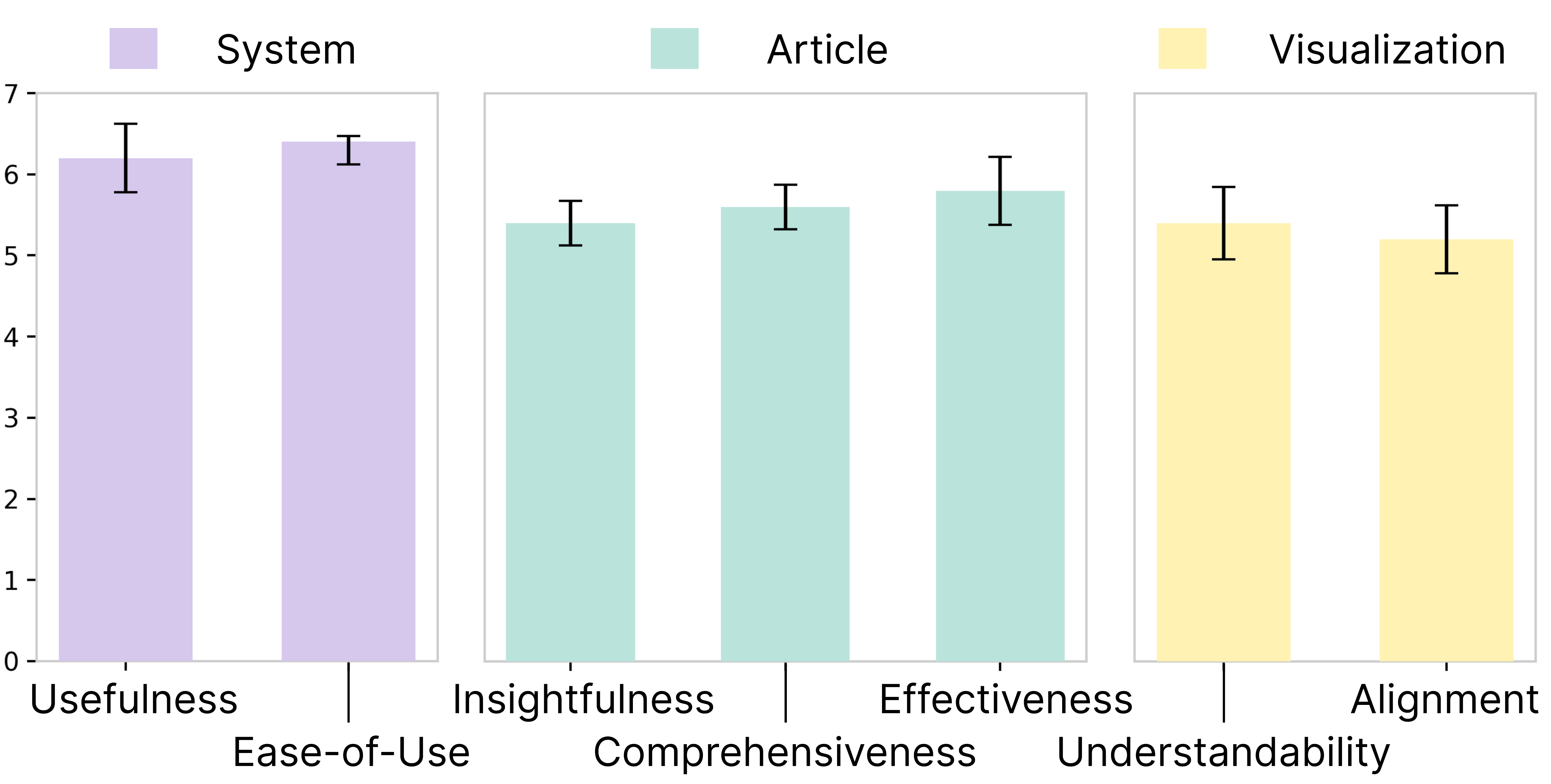}
  \caption{The ratings of system, article, and visualization from different criteria on a 7-point Likert scale (1=strongly disagree, 7=strongly agree).}
  \label{fig:expert_questionaire}
  \vspace{-1em}
\end{figure}

\textbf{Results.} Fig.~\ref{fig:expert_questionaire} shows the questionnaire results. Generally, all experts agree that the automated data-driven article composition system is a useful and easy-to-use tool, with articles effectively fulfilling the requirements and understandable visualizations.

\textit{\underline{Quality of decision-making.}} To quantitatively evaluate the alignment between \name's selections and expert preferences, we analyzed element selection congruence across the narrative composition space. For each dimension, we identified the \textit{modal expert choice} (most selected element in baseline trials). \name achieved 59\% congruence with these expert-preferred elements overall, demonstrating its capacity to replicate domain experts' collective judgment. Notably, congruence varied by dimension: Narrative Perspective (100\%), Narrative Structure (80\%), and Narrative Intent (80\%) showed strong alignment, while Headline (40\%) and Narrative Pattern (35.7\%) exhibited lower matches. The lower congruence in dimensions may stem from their greater subjectivity and stylistic variability, while higher-alignment dimensions often involve more functional or goal-driven decisions, which are easier for \name to model and generalize. Despite the mismatches, all experts found the system’s selections reasonable and well-justified. For instance, E1 noted that “although I initially preferred third-person narration, the system’s choice of second-person narration resulted in a more immersive and engaging tone”.

\textit{\underline{Quality of generated articles.}} Most experts (E1-E2, E4-E5) agreed the generated articles effectively fulfilled requirements. E2 highlighted that the content strongly matched the requirements and selected design elements, accompanied by clear visualizations. However, four experts pointed out that the writing style felt slightly "AI-generated," with limited naturalness. These findings suggest that while the content met core requirements, refining the language style could improve readability and engagement. Overall, most experts (E1, E3-E5) viewed the articles as useful prototypes for idea structuring and believed that improvements in tone and personalization could further enhance their quality.

\textit{\underline{Usability of system.}} All experts were satisfied with the clear system interface and agreed it effectively supported article creation. E2 noted that the system could significantly improve work efficiency by supporting complementary strengths in writing and visualization. E4 highlighted that faster extract insight extraction reduced time spent on data analysis. She also suggested that ranking insights by importance could improve clarity. Furthermore, she appreciated the design space view, which inspired new ideas and enabled rapid exploration of different styles. E1 also praised the chain-of-thought view, saying that “seeing the system’s running process increased my trust in the system”.

\begin{table}[t]
\centering
\begin{tabular}{l p{5.4cm}}
\hline
\textbf{Design Dimensions} & \textbf{Design Elements} \\
\hline
\cellcolor{dimgray}\textbf{Non-numerical Field} & Categorical or temporal fields, e.g. \textit{city}, \textit{date} \\
\cellcolor{dimgray}\textbf{Numerical Field} & Numerical fields with proportional or quantitative values, e.g. \textit{rate}, \textit{cost} \\
\cellcolor{dimgray}\textbf{Binding Type} & Semantics (with color) to category, Area to quantity, Unit to quantity, Unit to proportion, Unit filled with color to proportion \\
\cellcolor{dimgray}\textbf{Chart Type} & Block chart, Bar chart, Line chart, Pie chart, Donut chart, Number chart \\
\cellcolor{dimgray}\textbf{Icon Strategy} & Same icon, Different icons \\
\cellcolor{dimgray}\textbf{Icon Theme} & Food+Beverage, Shapes, Numbers, Alphabet, Animals, Sports+Fitness, Transportation, Genders, … (and more) \\
\cellcolor{dimgray}\textbf{Background Color} & \multirow{3}{5.5cm}{Red, Magenta, Purple, Blue purple, Blue, Cyan, Green, Lime, Yellow, Amber, Orange, Red Orange, Gray} \\
\cellcolor{dimgray}\textbf{Description Color} & \\
\cellcolor{dimgray}\textbf{Field Color} & \\
\hline
\end{tabular}
\vspace{0.5em}
\caption{Overview of the pictorial visualization space.}
\label{tab:picvis}
\vspace{-1em}
\end{table}

\section{Case II: Pictorial Visualization}
To further validate \name’s effectiveness and cross-domain applicability, we apply it to pictorial visualization generation. Pictorial visualizations employ iconic metaphors to transform abstract data into intuitive representations through decisions on data binding, chart types, icon semantics, and color harmony. We define a structured design space that organizes these decisions and implements \name to navigate this space, followed by an evaluation showcasing generated examples and a user study assessing quality.

\subsection{Design Space of Pictorial Visualization}
Synthesizing and extending foundational dimensions from Text-to-Viz~\cite{cui2019text} and Vistylist~\cite{shi2022supporting}, we formalize the pictorial visualization design space into nine core dimensions: 

\begin{itemize}
    \item Non-numerical Field: Specifies categorical or temporal attributes used for semantic grouping.
    \item Numerical Field: Defines quantitative values to visualize, such as prices, counts, or percentages.
    \item Binding Type: Governs how data maps to visual properties through specific encoding methods.
    \item Chart Type:  Specifies the chart used to visualize data.
    \item Icon Strategy: Controls whether categories use the same or different icons.
    \item Icon Theme: Aligns icons with 48 domain contexts.
    \item Background Color: Sets the canvas foundation color, prioritizing high contrast with foreground elements
    \item Description Color: Defines the color for textual annotations that explain visualized data points.
    \item Field Color: Sets domain-aligned palettes for data fields, ensuring semantic and color harmony.
\end{itemize}

Table~\ref{tab:picvis} lists dimensions and elements. Notably, the field color dimension is dynamically instantiated based on categorical field values. For instance, an animal field containing cat and dog would generate \textit{field color cat} and \textit{field color dog} as distinct dimensions, which ensures semantic relevance when assigning domain-specific colors.

\begin{figure*}[th]
  \centering
  \includegraphics[width=0.95\textwidth]{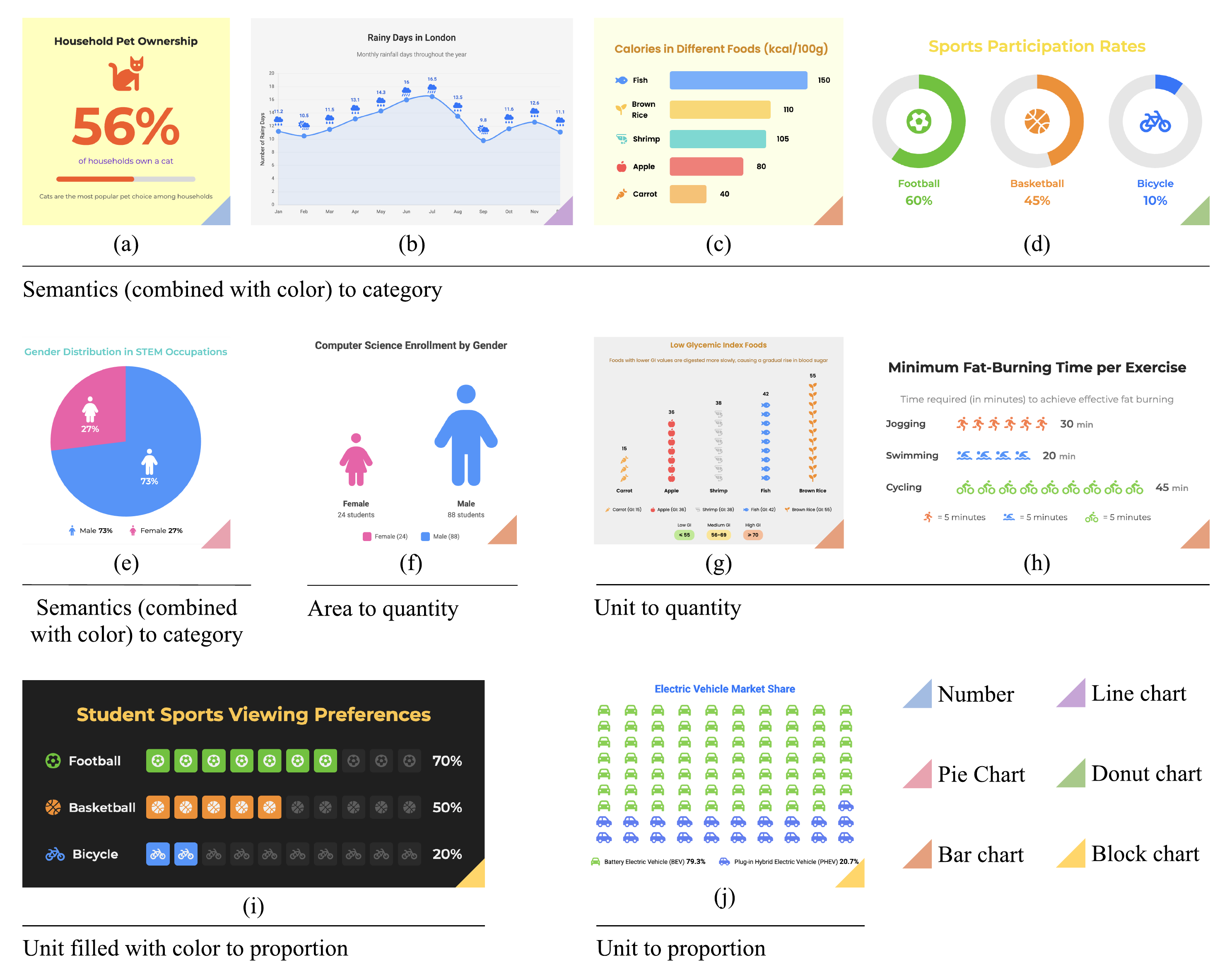}
  \caption{Examples generated by \name based on the following requirements: (a) "A clear, visually focused design using a cheerful and playful color palette highlighting a single important percentage: the household ownership rate of cats"; (b) "A pictorial visualization that reflects the variation in the number of rainy days in London over the course of 12 months"; (c) "A pictorial visualization that ranks foods by their calories"; (d) "A clean circular representation where each sport is shown with its own individual percentage of total student participation"; (e) "A pictorial visualization comparing the proportion of males and females working in STEM fields"; (f) "A pictorial visualization that shows male and female enrollment in computer science using proportional icon sizes"; (g) "A pictorial visualization that vividly showcases low GI foods"; (h) "A pictorial visualization that compares the fat-burning efficiency of different exercises"; (i) "A pictorial visualization with dark-themed style where each sport has its own proportion shown, to illustrate student viewing preferences"; (j) "A pictorial visualization that highlights the market share of BEV relative to PHEV sales".}
  \label{fig:picvis}
  \vspace{-1em}
\end{figure*}

\subsection{Implementation Details}
\name accepts a user requirement and a dataset as input. It first utilizes DeepSeek-R1~\cite{guo2025deepseek} to generate design constraints based on the input and design space, then searches the solution using MCTS. Finally, the \texttt{generate\_pictorial\_vis} action function invokes Claude 3.7 Sonnet~\cite{claude_3.7} to synthesize HTML code that implements the optimal design solution, leveraging Font Awesome~\cite{fontawesome} icons aligned with the selected icon theme to ensure aesthetic integrity.

\subsection{Evaluation}
We first showcase example results generated by \name to highlight the expressiveness of the designs. We then conduct a questionnaire-based user study to evaluate design quality.

\subsubsection{Example Results}
We present a variety of pictorial visualizations generated by \name in Fig.~\ref{fig:picvis} that illustrate five data binding types and six chart types. For example, Fig.~\ref{fig:picvis}(a) uses a large numeric value with a cat icon and progress bar on a yellow background to vividly emphasize high cat ownership rates. Fig.~\ref{fig:picvis}(b) employs a line chart to visualize temporal field data, where rainfall duration is encoded through raindrop icons to indicate different precipitation days. Fig.~\ref{fig:picvis}(f) selects area-to-quantity binding with scaled gender icons proportional to enrollment numbers. Fig.~\ref{fig:picvis}(h) utilizes horizontal bars stacking sport-icon units to contrast fat-burning rates through visual density. For proportional data, Fig.~\ref{fig:picvis}(i) colors partial units of sports icons to denote student viewing preferences. When proportions sum to one, pie charts (Fig.~\ref{fig:picvis}(e)) and block charts (Fig.~\ref{fig:picvis}(j)) are used to effectively convey part-to-whole relationships through distinct colors and representative icons. These examples showcase \name's ability to translate abstract data into intuitive visual metaphors based on user requirements.

\subsubsection{Evaluation of the Generated Designs}
We evaluate \name's effectiveness in generating high-quality designs through a questionnaire-based user study comparing pictorial visualizations created by \name and those generated by a large language model.

\textbf{Baseline.} We adopt a baseline method using the same LLM (Claude 3.7 Sonnet) and design space as \name. Driven by a single prompt, the model takes the user requirement and dataset as input, selects design elements, and generates the pictorial visualization. For a fair comparison, the prompt mirrors the constraint generation prompt of \name, excluding only the constraint semantics to guide element selection. The final HTML code is synthesized using the same generation rules as \name to ensure output consistency.

\textbf{Materials.} We collected 10 datasets and defined a specific user requirement for each to guide the visualization. Using these inputs, two pictorial visualizations were generated per dataset, one using \name and one using the baseline, resulting in 10 groups and a total of 20 visualizations.

\textbf{Procedure.} We recruited 60 participants aged 20-32 ($M = 23.42$, $SD = 2.78$) from diverse backgrounds, such as art and design, computer science, and data science. All participants had experience in visualization: less than 1 year (26), 1–2 years (14), 2–5 years (18), and more than 5 years (2). In the questionnaire, participants reviewed 10 groups of visualizations, with two visualizations per group presented in random order. For each group, participants read the dataset and requirement, then chose the visualization that performed better in data accuracy, requirement consistency, and overall quality, or selected that both were effective.

\begin{figure}
  \centering
  \includegraphics[width=0.25\textwidth]{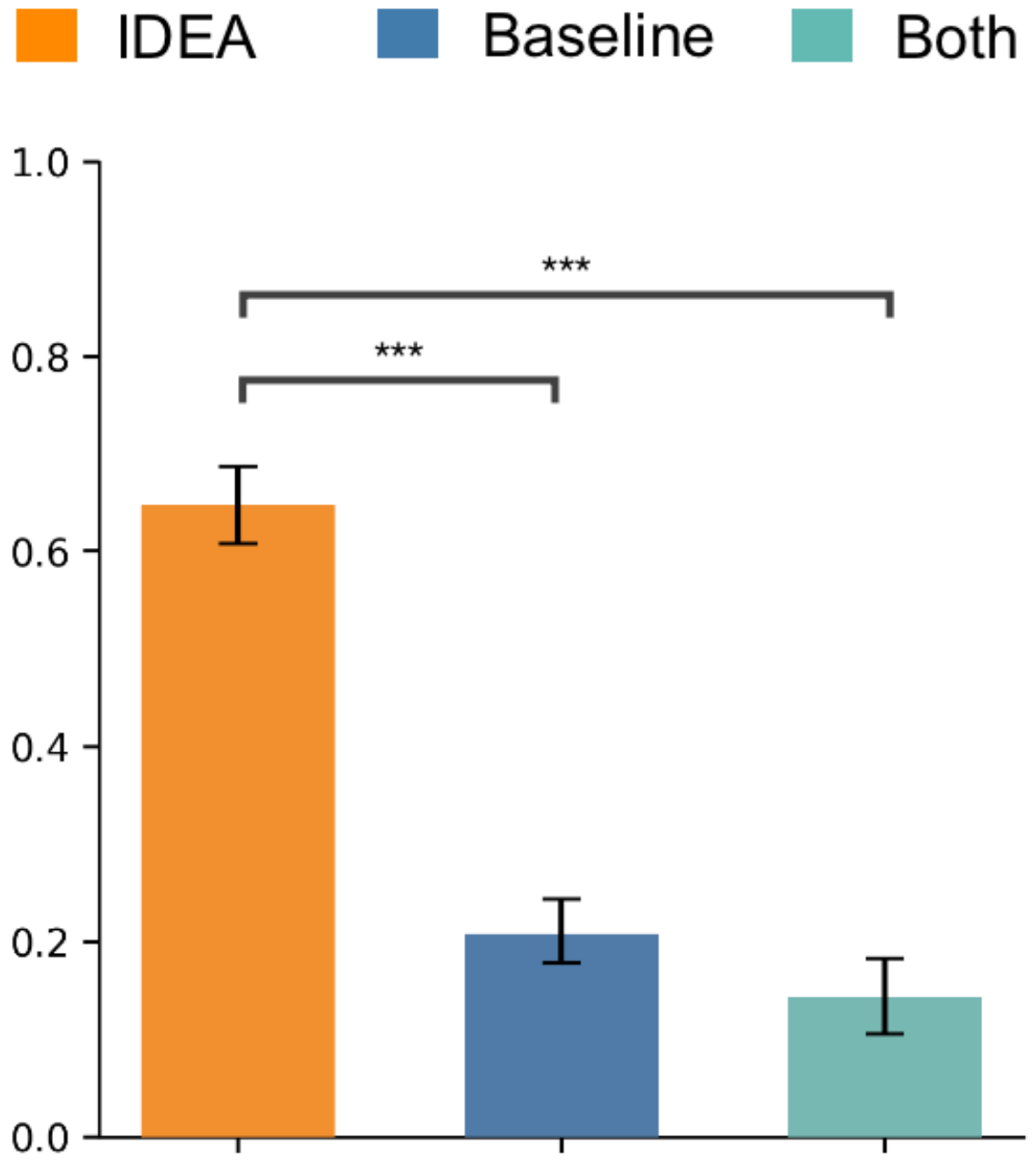}
  \caption{Participant selection proportions for \name, the baseline, and both options across 10 visualization groups. Horizontal brackets indicate pairwise significant difference ($*: p<.05, **: p<.01, ***: p<.001$). The error bars represent the $95\%$ confidence intervals.}
  \label{fig:pictorial_user_study}
\end{figure}

\textbf{Results.} We recorded the number of times each participant selected \name, the baseline, or both as effective across the 10 groups. The proportion of times was then analyzed using pairwise Wilcoxon signed-rank tests. As illustrated in Fig.~\ref{fig:pictorial_user_study}. \name ($M=0.648$, $SD=0.158$) was selected significantly more often than the baseline ($M = 0.208$, $SD = 0.127$) and both ($M = 0.143$, $SD = 0.162$), suggesting that \name produced higher-quality designs.

\section{Discussion}
This section begins by exploring potential application scenarios of \name in broader contexts. We then discuss its current limitations and outline directions for future work.

\subsection{Potential Application Scenarios}
\name’s ability to automate decision-making through constraint-guided search opens new possibilities for computational design across domains. For instance, \name can be applied to \textbf{interior design} by constructing a design space with dimensions like style, flooring materials, wall treatments, and layouts. It automatically generates personalized solutions based on user input, such as "Nordic-style living room," by applying constraints to exclude over-budget materials and prioritizing combinations like light-toned wood floors. It then translates the solution into structured prompts and generates photorealistic renderings using diffusion models~\cite{shao2024new,chen2023generating}, enabling rapid iteration and empowering non-expert homeowners to design independently. In \textbf{product design}, \name automates design by structuring spaces around core attributes like aesthetic style, color palette, and functionality. It executes the same pipeline to generate product renderings, reducing cognitive load for novice designers and streamlining the design process.

\subsection{Limitations and Future Work}
Despite the evaluations have demonstrated the effectiveness of \name, we summarize several limitations identified during implementation and evaluation to inform future research and encourage continued exploration.

\textit{\underline{Supporting Automated Design Space Construction.}} A key limitation of \name is its reliance on manually predefined design spaces, which require domain expertise and considerable time to synthesize knowledge. Future work could explore automated construction using LLMs to extract dimensions and elements from domain corpora, with an interactive interface enabling human-AI co-editing to refine the output and reduce setup time.

\textit{\underline{Enhancing Framework Usability.}} While \name supports rapid prototyping of intelligent design tools across domains, applying it to new scenarios still requires redefining action functions. Providing a set of predefined actions (e.g., text/image generation) and a visual interface for mapping dimensions to actions could simplify this process and enable non-programmers to build tools with minimal coding.

\textit{\underline{Improving Generation Quality.}} While evaluations indicate general satisfaction with Form's outcomes, experts noted some quality issues like overly artificial outputs and misaligned choices, which stem from the hallucinations of LLMs. As LLMs evolve, the generation quality will gradually improve. Approaches like integrating correction agents to validate constraints and refine outcomes could further enhance stability and reliability.

\textit{\underline{Dynamic Constraint-Search Coupling.}} \name's pre-generated constraints provide static guidance during solution search, which may restrict adaptability to emergent patterns during exploration. Future work could embed lightweight LLM agents within MCTS to dynamically refine constraints based on real-time search states.

\section{Conclusion}
We have proposed a structured design space model that formalizes design knowledge as machine-interpretable decision structures composed of orthogonal dimensions and discrete elements. Building on this model, we have introduced \name, a design decision-making framework that transforms user requirements into effective outcomes through LLM-based constraint generation, MCTS-guided design solution search, and domain-specific action execution. Through case studies in data-driven article composition and pictorial visualization generation, we demonstrated \name’s cross-domain applicability and validated its effectiveness through example results, expert interviews, and a user study. The evaluation confirmed \name’s ability to generate high-quality outcomes while supporting efficient and human-aligned decision-making. We envision expanding \name with automated design space construction, predefined actions, correction mechanisms, and dynamic constraint-search coupling to further advance intelligent design.

\section*{ACKNOWLEDGMENTS}
Nan Cao is the corresponding author. This work was supported by the National Key Research and Development Program of China (2023YFB3107100). We would like to thank all the reviewers for their valuable feedback.

\bibliographystyle{IEEEtran}
\bibliography{reference}

\begin{thebibliography}{10}
\providecommand{\url}[1]{#1}
\csname url@samestyle\endcsname
\providecommand{\newblock}{\relax}
\providecommand{\bibinfo}[2]{#2}
\providecommand{\BIBentrySTDinterwordspacing}{\spaceskip=0pt\relax}
\providecommand{\BIBentryALTinterwordstretchfactor}{4}
\providecommand{\BIBentryALTinterwordspacing}{\spaceskip=\fontdimen2\font plus
\BIBentryALTinterwordstretchfactor\fontdimen3\font minus \fontdimen4\font\relax}
\providecommand{\BIBforeignlanguage}[2]{{%
\expandafter\ifx\csname l@#1\endcsname\relax
\typeout{** WARNING: IEEEtran.bst: No hyphenation pattern has been}%
\typeout{** loaded for the language `#1'. Using the pattern for}%
\typeout{** the default language instead.}%
\else
\language=\csname l@#1\endcsname
\fi
#2}}
\providecommand{\BIBdecl}{\relax}
\BIBdecl

\bibitem{shu2020makes}
X.~Shu, A.~Wu, J.~Tang, B.~Bach, Y.~Wu, and H.~Qu, ``What makes a data-gif understandable?'' \emph{IEEE Transactions on Visualization and Computer Graphics}, vol.~27, no.~2, pp. 1492--1502, 2020.

\bibitem{halskov2021filtering}
K.~Halskov and C.~Lundqvist, ``Filtering and informing the design space: Towards design-space thinking,'' \emph{ACM Transactions on Computer-Human Interaction (TOCHI)}, vol.~28, no.~1, pp. 1--28, 2021.

\bibitem{chen2024viseval}
N.~Chen, Y.~Zhang, J.~Xu, K.~Ren, and Y.~Yang, ``Viseval: A benchmark for data visualization in the era of large language models,'' \emph{IEEE Transactions on Visualization and Computer Graphics}, 2024.

\bibitem{chen2022vizbelle}
Q.~Chen, Z.~Liu, C.~Wang, X.~Lan, Y.~Chen, S.~Chen, and N.~Cao, ``Vizbelle: A design space of embellishments for data visualization,'' \emph{arXiv preprint arXiv:2209.03642}, 2022.

\bibitem{lan2021kineticharts}
X.~Lan, Y.~Shi, Y.~Wu, X.~Jiao, and N.~Cao, ``Kineticharts: Augmenting affective expressiveness of charts in data stories with animation design,'' \emph{IEEE Transactions on Visualization and Computer Graphics}, vol.~28, no.~1, pp. 933--943, 2021.

\bibitem{cui2019text}
W.~Cui, X.~Zhang, Y.~Wang, H.~Huang, B.~Chen, L.~Fang, H.~Zhang, J.-G. Lou, and D.~Zhang, ``Text-to-viz: Automatic generation of infographics from proportion-related natural language statements,'' \emph{IEEE transactions on visualization and computer graphics}, vol.~26, no.~1, pp. 906--916, 2019.

\bibitem{zhu2021augmenting}
C.~Zhu-Tian, S.~Ye, X.~Chu, H.~Xia, H.~Zhang, H.~Qu, and Y.~Wu, ``Augmenting sports videos with viscommentator,'' \emph{IEEE Transactions on Visualization and Computer Graphics}, vol.~28, no.~1, pp. 824--834, 2021.

\bibitem{mackinlay1986automating}
J.~Mackinlay, ``Automating the design of graphical presentations of relational information,'' \emph{Acm Transactions On Graphics (Tog)}, vol.~5, no.~2, pp. 110--141, 1986.

\bibitem{shaw2011role}
M.~Shaw, ``The role of design spaces,'' \emph{IEEE software}, vol.~29, no.~1, pp. 46--50, 2011.

\bibitem{schulz2010design}
H.-J. Schulz, S.~Hadlak, and H.~Schumann, ``The design space of implicit hierarchy visualization: A survey,'' \emph{IEEE transactions on visualization and computer graphics}, vol.~17, no.~4, pp. 393--411, 2010.

\bibitem{sarikaya2017scatterplots}
A.~Sarikaya and M.~Gleicher, ``Scatterplots: Tasks, data, and designs,'' \emph{IEEE transactions on visualization and computer graphics}, vol.~24, no.~1, pp. 402--412, 2017.

\bibitem{hografer2020map}
M.~Hogr{\"a}fer, M.~Heitzler, and H.-J. Schulz, ``The state of the art in map-like visualization,'' in \emph{Computer Graphics Forum}, vol.~39, no.~3.\hskip 1em plus 0.5em minus 0.4em\relax Wiley Online Library, 2020, pp. 647--674.

\bibitem{bludau2023unfolding}
M.-J. Bludau, M.~D{\"o}rk, and C.~Tominski, ``Unfolding edges: Adding context to edges in multivariate graph visualization,'' in \emph{Computer Graphics Forum}, vol.~42, no.~3.\hskip 1em plus 0.5em minus 0.4em\relax Wiley Online Library, 2023, pp. 297--309.

\bibitem{nobre2019state}
C.~Nobre, M.~Meyer, M.~Streit, and A.~Lex, ``The state of the art in visualizing multivariate networks,'' in \emph{Computer Graphics Forum}, vol.~38, no.~3.\hskip 1em plus 0.5em minus 0.4em\relax Wiley Online Library, 2019, pp. 807--832.

\bibitem{rufiange2012treematrix}
S.~Rufiange, M.~J. McGuffin, and C.~P. Fuhrman, ``Treematrix: A hybrid visualization of compound graphs,'' in \emph{Computer Graphics Forum}, vol.~31, no.~1.\hskip 1em plus 0.5em minus 0.4em\relax Wiley Online Library, 2012, pp. 89--101.

\bibitem{bach2022dashboard}
B.~Bach, E.~Freeman, A.~Abdul-Rahman, C.~Turkay, S.~Khan, Y.~Fan, and M.~Chen, ``Dashboard design patterns,'' \emph{IEEE transactions on visualization and computer graphics}, vol.~29, no.~1, pp. 342--352, 2022.

\bibitem{yang2021design}
L.~Yang, X.~Xu, X.~Lan, Z.~Liu, S.~Guo, Y.~Shi, H.~Qu, and N.~Cao, ``A design space for applying the freytag's pyramid structure to data stories,'' \emph{IEEE Transactions on Visualization and Computer Graphics}, vol.~28, no.~1, pp. 922--932, 2021.

\bibitem{freytag1895technique}
G.~Freytag, \emph{Technique of the drama: An exposition of dramatic composition and art}.\hskip 1em plus 0.5em minus 0.4em\relax S. Griggs, 1895.

\bibitem{zhao2018characterizes}
N.~Zhao, Y.~Cao, and R.~W. Lau, ``What characterizes personalities of graphic designs?'' \emph{ACM Transactions on Graphics (TOG)}, vol.~37, no.~4, pp. 1--15, 2018.

\bibitem{baechler2024screenai}
G.~Baechler, S.~Sunkara, M.~Wang, F.~Zubach, H.~Mansoor, V.~Etter, V.~C{\u{a}}rbune, J.~Lin, J.~Chen, and A.~Sharma, ``Screenai: A vision-language model for ui and infographics understanding,'' \emph{arXiv preprint arXiv:2402.04615}, 2024.

\bibitem{yang2016automatic}
X.~Yang, T.~Mei, Y.-Q. Xu, Y.~Rui, and S.~Li, ``Automatic generation of visual-textual presentation layout,'' \emph{ACM Transactions on Multimedia Computing, Communications, and Applications (TOMM)}, vol.~12, no.~2, pp. 1--22, 2016.

\bibitem{chen2025posta}
H.~Chen, X.~Xu, W.~Li, J.~Ren, T.~Ye, S.~Liu, Y.-C. Chen, L.~Zhu, and X.~Wang, ``Posta: A go-to framework for customized artistic poster generation,'' \emph{arXiv preprint arXiv:2503.14908}, 2025.

\bibitem{wang2025designdiffusion}
Z.~Wang, J.~Bao, S.~Gu, D.~Chen, W.~Zhou, and H.~Li, ``Designdiffusion: High-quality text-to-design image generation with diffusion models,'' \emph{arXiv preprint arXiv:2503.01645}, 2025.

\bibitem{rokach2005decision}
L.~Rokach and O.~Maimon, ``Decision trees,'' \emph{Data mining and knowledge discovery handbook}, pp. 165--192, 2005.

\bibitem{goldberg1989genetic}
D.~E. Goldberg, ``Genetic algorithm in search, optimization and machine learning, addison,'' \emph{W esley Publishing Company, R eading, MA}, vol.~1, no.~98, p.~9, 1989.

\bibitem{sutton1998reinforcement}
R.~S. Sutton, A.~G. Barto \emph{et~al.}, \emph{Reinforcement learning: An introduction}.\hskip 1em plus 0.5em minus 0.4em\relax MIT press Cambridge, 1998, vol.~1, no.~1.

\bibitem{puterman2014markov}
M.~L. Puterman, \emph{Markov decision processes: discrete stochastic dynamic programming}.\hskip 1em plus 0.5em minus 0.4em\relax John Wiley \& Sons, 2014.

\bibitem{watkins1992q}
C.~J. Watkins and P.~Dayan, ``Q-learning,'' \emph{Machine learning}, vol.~8, pp. 279--292, 1992.

\bibitem{mnih2015human}
V.~Mnih, K.~Kavukcuoglu, D.~Silver, A.~A. Rusu, J.~Veness, M.~G. Bellemare, A.~Graves, M.~Riedmiller, A.~K. Fidjeland, G.~Ostrovski \emph{et~al.}, ``Human-level control through deep reinforcement learning,'' \emph{nature}, vol. 518, no. 7540, pp. 529--533, 2015.

\bibitem{williams1992simple}
R.~J. Williams, ``Simple statistical gradient-following algorithms for connectionist reinforcement learning,'' \emph{Machine learning}, vol.~8, pp. 229--256, 1992.

\bibitem{schulman2017proximal}
J.~Schulman, F.~Wolski, P.~Dhariwal, A.~Radford, and O.~Klimov, ``Proximal policy optimization algorithms,'' \emph{arXiv preprint arXiv:1707.06347}, 2017.

\bibitem{browne2012survey}
C.~B. Browne, E.~Powley, D.~Whitehouse, S.~M. Lucas, P.~I. Cowling, P.~Rohlfshagen, S.~Tavener, D.~Perez, S.~Samothrakis, and S.~Colton, ``A survey of monte carlo tree search methods,'' \emph{IEEE Transactions on Computational Intelligence and AI in games}, vol.~4, no.~1, pp. 1--43, 2012.

\bibitem{cao2024survey}
Y.~Cao, H.~Zhao, Y.~Cheng, T.~Shu, Y.~Chen, G.~Liu, G.~Liang, J.~Zhao, J.~Yan, and Y.~Li, ``Survey on large language model-enhanced reinforcement learning: Concept, taxonomy, and methods,'' \emph{IEEE Transactions on Neural Networks and Learning Systems}, 2024.

\bibitem{yang2024selfgoal}
R.~Yang, J.~Chen, Y.~Zhang, S.~Yuan, A.~Chen, K.~Richardson, Y.~Xiao, and D.~Yang, ``Selfgoal: Your language agents already know how to achieve high-level goals,'' in \emph{NeurIPS 2024 Workshop on Open-World Agents}, 2024.

\bibitem{tan2024true}
W.~Tan, W.~Zhang, S.~Liu, L.~Zheng, X.~Wang, and B.~An, ``True knowledge comes from practice: Aligning llms with embodied environments via reinforcement learning,'' \emph{arXiv preprint arXiv:2401.14151}, 2024.

\bibitem{ahn2022can}
M.~Ahn, A.~Brohan, N.~Brown, Y.~Chebotar, O.~Cortes, B.~David, C.~Finn, C.~Fu, K.~Gopalakrishnan, K.~Hausman \emph{et~al.}, ``Do as i can, not as i say: Grounding language in robotic affordances,'' \emph{arXiv preprint arXiv:2204.01691}, 2022.

\bibitem{yan2023ask}
X.~Yan, Y.~Song, X.~Cui, F.~Christianos, H.~Zhang, D.~H. Mguni, and J.~Wang, ``Ask more, know better: Reinforce-learned prompt questions for decision making with large language models,'' \emph{arXiv preprint arXiv:2310.18127}, 2023.

\bibitem{xie2024haichart}
Y.~Xie, Y.~Luo, G.~Li, and N.~Tang, ``Haichart: Human and ai paired visualization system,'' \emph{arXiv preprint arXiv:2406.11033}, 2024.

\bibitem{shi2020calliope}
D.~Shi, X.~Xu, F.~Sun, Y.~Shi, and N.~Cao, ``Calliope: Automatic visual data story generation from a spreadsheet,'' \emph{IEEE Transactions on Visualization and Computer Graphics}, vol.~27, no.~2, pp. 453--463, 2020.

\bibitem{shen2024data}
L.~Shen, H.~Li, Y.~Wang, and H.~Qu, ``From data to story: Towards automatic animated data video creation with llm-based multi-agent systems,'' in \emph{2024 IEEE VIS Workshop on Data Storytelling in an Era of Generative AI (GEN4DS)}.\hskip 1em plus 0.5em minus 0.4em\relax IEEE, 2024, pp. 20--27.

\bibitem{hao2024design}
S.~Hao, Z.~Wang, B.~Bach, and L.~Pschetz, ``Design patterns for data-driven news articles,'' in \emph{Proceedings of the 2024 CHI Conference on Human Factors in Computing Systems}, 2024, pp. 1--16.

\bibitem{bach2018narrative}
B.~Bach, M.~Stefaner, J.~Boy, S.~Drucker, L.~Bartram, J.~Wood, P.~Ciuccarelli, Y.~Engelhardt, U.~Koeppen, and B.~Tversky, ``Narrative design patterns for data-driven storytelling,'' in \emph{Data-driven storytelling}.\hskip 1em plus 0.5em minus 0.4em\relax AK Peters/CRC Press, 2018, pp. 107--133.

\bibitem{ojo2018patterns}
A.~Ojo and B.~Heravi, ``Patterns in award winning data storytelling: Story types, enabling tools and competences,'' \emph{Digital journalism}, vol.~6, no.~6, pp. 693--718, 2018.

\bibitem{wikinarrative}
\BIBentryALTinterwordspacing
W.~contributors. (2025) Narration. [Online]. Available: \url{https://en.wikipedia.org/wiki/Narration}
\BIBentrySTDinterwordspacing

\bibitem{lifschitz2019answer}
V.~Lifschitz, \emph{Answer set programming}.\hskip 1em plus 0.5em minus 0.4em\relax Springer Cham, 2019, vol.~3.

\bibitem{gebser2014clingo}
M.~Gebser, R.~Kaminski, B.~Kaufmann, and T.~Schaub, ``Clingo= asp+ control: Preliminary report,'' \emph{arXiv preprint arXiv:1405.3694}, 2014.

\bibitem{kocsis2006bandit}
L.~Kocsis and C.~Szepesv{\'a}ri, ``Bandit based monte-carlo planning,'' in \emph{European conference on machine learning}.\hskip 1em plus 0.5em minus 0.4em\relax Springer, 2006, pp. 282--293.

\bibitem{guo2025deepseek}
D.~Guo, D.~Yang, H.~Zhang, J.~Song, R.~Zhang, R.~Xu, Q.~Zhu, S.~Ma, P.~Wang, X.~Bi \emph{et~al.}, ``Deepseek-r1: Incentivizing reasoning capability in llms via reinforcement learning,'' \emph{arXiv preprint arXiv:2501.12948}, 2025.

\bibitem{shi2022supporting}
Y.~Shi, P.~Liu, S.~Chen, M.~Sun, and N.~Cao, ``Supporting expressive and faithful pictorial visualization design with visual style transfer,'' \emph{IEEE Transactions on Visualization and Computer Graphics}, vol.~29, no.~1, pp. 236--246, 2022.

\bibitem{claude_3.7}
\BIBentryALTinterwordspacing
{Anthropic}, ``Claude 3.7 sonnet,'' 2025, large language model developed by Anthropic. [Online]. Available: \url{https://www.anthropic.com/news/claude-3-7-sonnet}
\BIBentrySTDinterwordspacing

\bibitem{fontawesome}
\BIBentryALTinterwordspacing
F.~Awesome. (2025) Font awesome. [Online]. Available: \url{https://fontawesome.com/}
\BIBentrySTDinterwordspacing

\bibitem{shao2024new}
Z.~Shao, J.~Chen, H.~Zeng, W.~Hu, Q.~Xu, and Y.~Zhang, ``A new approach to interior design: Generating creative interior design videos of various design styles from indoor texture-free 3d models,'' \emph{Buildings}, vol.~14, no.~6, p. 1528, 2024.

\bibitem{chen2023generating}
J.~Chen, Z.~Shao, and B.~Hu, ``Generating interior design from text: A new diffusion model-based method for efficient creative design,'' \emph{Buildings}, vol.~13, no.~7, p. 1861, 2023.

\end{thebibliography}





\begin{IEEEbiography}[{\includegraphics[width=1in,height=1.25in,clip,keepaspectratio]{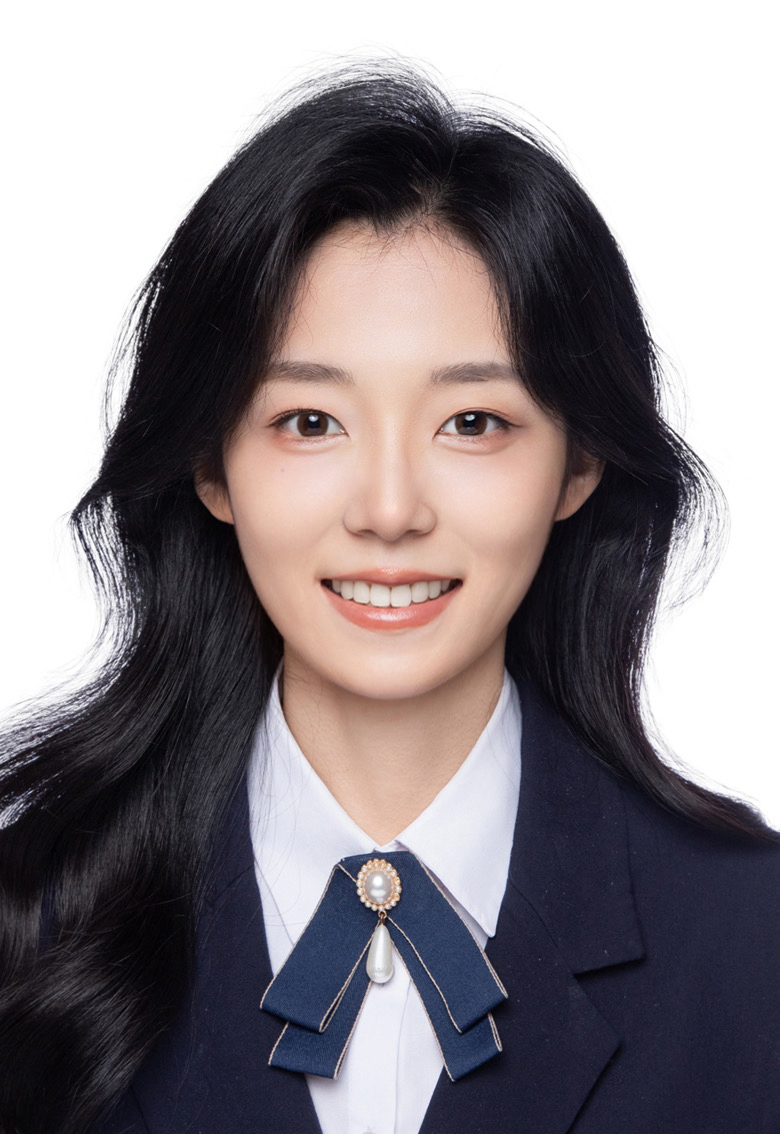}}]{Chuer Chen}
received her MSc degree from the Department of Electrical and Computer Engineering, National University of Singapore in 2021. She is currently working toward her Ph.D. degree as part of the Intelligent Big Data Visualization (iDVx) Lab, Tongji University. Her research interests include information visualization and intelligent design.
\end{IEEEbiography}

\begin{IEEEbiography}[{\includegraphics[width=1in,height=1.25in,clip,keepaspectratio]{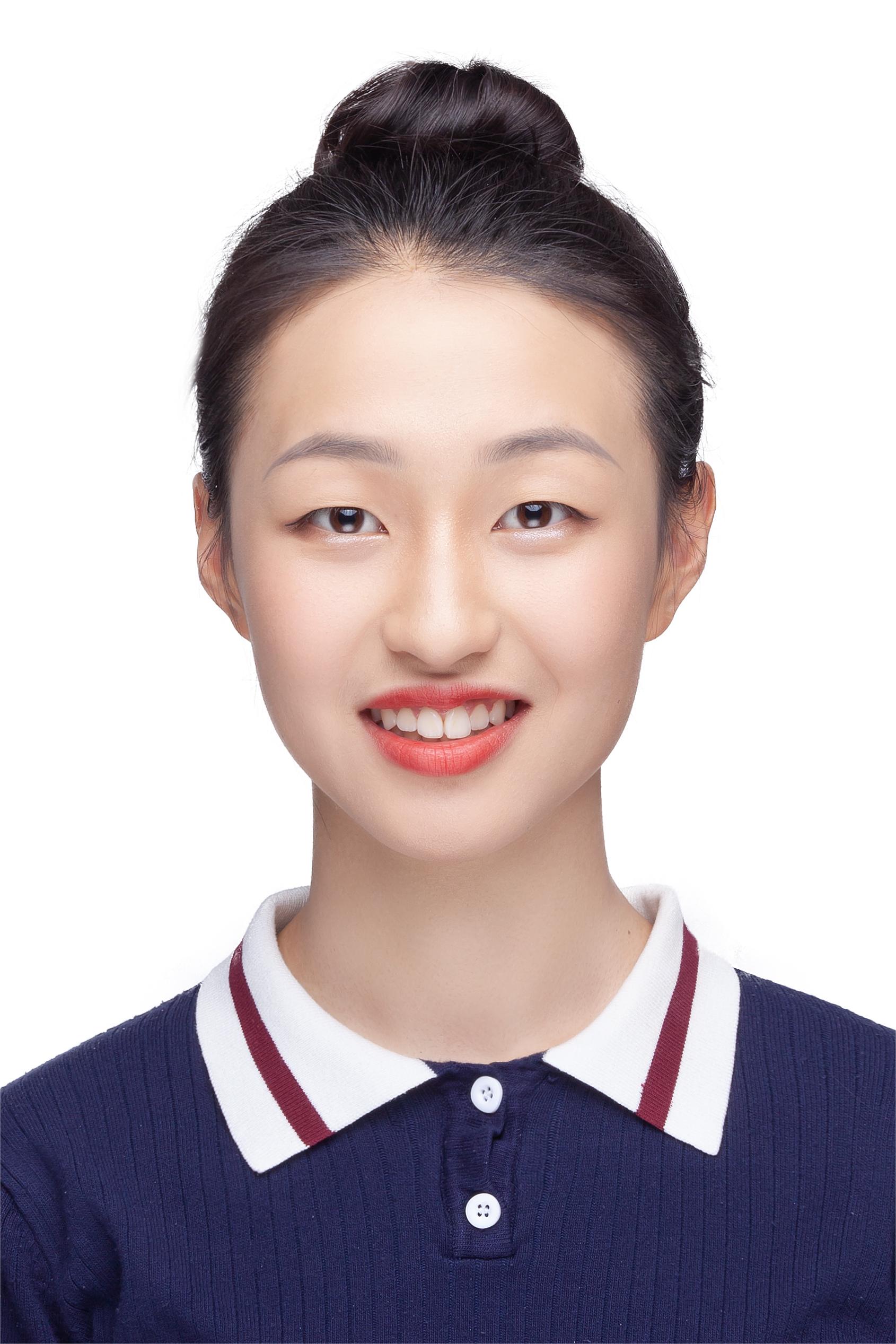}}]{Xiaoke Yan}
 received her bachelor’s degree from the Department of Software Engineering, Tongji University in 2024. She is currently pursuing her master’s degree at the College of Design and Innovation, Tongji University, where she is a member of the Intelligent Big Data Visualization Lab (iDVx Lab). Her research interests include information visualization, AI-supported design, and human-computer interaction.
\end{IEEEbiography}

\begin{IEEEbiography}[{\includegraphics[width=1in,height=1.25in,clip,keepaspectratio]{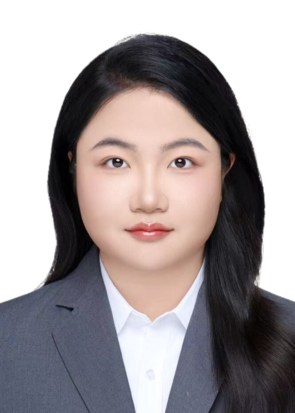}}]{Xiaoyu Qi}
received her bachelor's degree from the Department of Software Engineering, Tongji University in 2022. Currently, she is a master's candidate at Tongji University. Her research interests include AI-supported design and data visualization.
\end{IEEEbiography}

\begin{IEEEbiography}[{\includegraphics[width=1in,height=1.25in,clip,keepaspectratio]{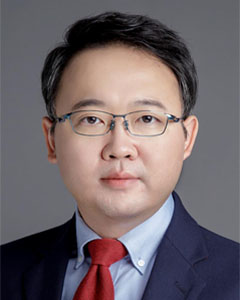}}]{Nan Cao}
received his Ph.D. degree in Computer Science and Engineering from the Hong Kong University of Science and Technology (HKUST), Hong Kong, China in 2012. He is currently a professor at Tongji University and the Vice Dean of the Tongji College of Design and Innovation. He also directs the Tongji Intelligent Big Data Visualization Lab (iDV$^x$ Lab) and conducts interdisciplinary research across multiple fields, including data visualization, human-computer interaction, and artificial intelligence. He was a research staff member at the IBM T.J. Watson Research Center, New York, NY, USA before joining the Tongji faculty in 2016.
\end{IEEEbiography}

\end{document}